# Free Energy Landscape of Colloidal Clusters in Spherical Confinement

*Junwei Wang[1], Chrameh Fru Mbah[2], Thomas Przybilla[3], Silvan Englisch[3], Erdmann Spiecker[3], Michael Engel[2,*], Nicolas Vogel[1*]*

1 Institute of Particle Technology,

2 Institute for Multiscale Simulation, IZNF,

3 Institute of Micro- and Nanostructure Research (IMN) & Center for Nanoanalysis and Electron Microscopy (CENEM), IZNF,

Friedrich-Alexander University Erlangen-Nürnberg, 91058 Erlangen, Germany.

* Corresponding Authors: michael.engel@fau.de, nicolas.vogel@fau.de

# ABSTRACT

The structure of finite self-assembling systems depends sensitively on the number of constituent building blocks. Recently, it was demonstrated that hard sphere-like colloidal particles show a magic number effect when confined in spherical emulsion droplets. Geometric construction rules permit a few dozen magic numbers that correspond to a discrete series of completely filled concentric icosahedral shells. Here, we investigate the free energy landscape of these colloidal clusters as a function of the number of their constituent building blocks for system sizes up to several thousand particles. We find that minima in the free energy landscape, arising from the presence of filled, concentric shells, are significantly broadened. In contrast to their atomic analogues, colloidal clusters in spherical confinement can flexibly accommodate excess colloids by ordering icosahedrally in the cluster center while changing the structure near the cluster surface. In-between these magic number regions, the building blocks cannot arrange into filled shells. Instead, we observe that defects accumulate in a single wedge and therefore only affect a few tetrahedral grains of the cluster. We predict the existence of this wedge by simulation and confirm its presence in experiment using electron tomography. The introduction of the wedge minimizes the free energy penalty by confining defects to small regions within the cluster. In addition, the remaining ordered tetrahedral grains can relax internal strain by breaking icosahedral symmetry. Our findings demonstrate how multiple defect mechanisms collude to form the complex free energy landscape of hard sphere-like colloidal clusters.

The number of particles is a crucial parameter to determine structure in a finite self-assembling system. Structure, on the other hand, governs physical properties. Thus, understanding the influence of system size on the formed structure is important to predict and control properties of discrete systems.

An important manifestation of this finite number effect is observed in atomic clusters – small aggregates of atoms, too small to be considered a bulk crystal, yet too large to be a molecule. Mass spectra of synthesized clusters often show a rapid variation of cluster size probability instead of a monotonous statistical distribution.[1] Clusters with comparably high occurrence appear due to enhanced thermodynamic stability and are termed magic number clusters.[1] Although the exact value of magic numbers depends on interatomic interactions,[2,3] certain magic numbers are found almost universally in atomic clusters. These numbers, especially 13, 55 and 147,[1,4] coincide with the number of spheres required to build a Mackay icosahedron, a geometric model proposed in 1962.[5] The Mackay model is a dense packing of equal spheres in form of concentric icosahedral shells around a central sphere. The associated magic numbers relate to the number of atoms permitting closed shells,[6] 13 atoms for one shell, 55 atoms for two shells, *etc*. Atomic clusters with Mackay icosahedral structures can be observed for larger system sizes, up to a few thousand atoms.[7,8] The icosahedral arrangement in such magic number clusters allows their surfaces to be covered exclusively by {111} crystal planes, in which each atom maximizes the number of its nearest neighbors and therefore minimizes its potential energy.[9] For system sizes between magic numbers, other cluster symmetries or irregular structures are competing.[10,11] During formation, atomic clusters can adjust the number of their atoms by adding or removing surface atoms to reach nearby magic numbers. As each system size is unique in its own, even a few more or less atoms change the system's thermodynamic properties, it is challenging to determine the number–structure relationship in atomic clusters of arbitrary sizes.[4,12]

Recently, magic number clusters were shown to exist in colloidal particles where the building blocks are of much larger size.[13] Interactions between charge-stabilized colloidal particles are significantly weaker compared to interactions of their atomic analogues, and structure formation can be accurately reproduced using hard sphere models.[13,14] In the absence of attractive interactions, colloidal cluster formation requires external confinement, for example by forming an emulsion droplet of an aqueous colloidal dispersion, which is subsequently dried.[13–26] In such confined self-assembling systems, the free energy of the system is governed by entropy alone.[14,27–29]

Similar to the case of atomic clusters, minimization of free energy, realized by maximization of entropy, dictates the structure of colloidal clusters. For up to several tens of thousands of confined colloidal particles, icosahedral symmetry or small variations thereof is preferred over face-centered cubic (fcc) arrangement, the latter being the most stable crystal structure in bulk.[14] This is rationalized by the geometry of the icosahedron. As the highest three-dimensional point group symmetry, icosahedral order naturally appears in locally dense packings near the curved interface of spherical droplets.[14,30] The spherical interface templates 20 ordered domains in icosahedral arrangement[31], which subsequently crystallize towards the center.[14] Free energy calculations show that colloidal clusters reach a minimum in the free energy landscape if and only if the numbers of particles are such that the system can form complete icosahedral concentric shells. As the number of particles increases, free energy cycles through successive minima, each of them corresponding to a fixed number of concentric shells.[13]

While the concept of magic number clusters bridges the atomic and the colloidal realm, there are differences. Structurally, atomic icosahedral clusters are usually described by the Mackay model. In contrast, in colloidal clusters formed through droplet confinement, the sharp Mackay icosahedral vertices are not well compatible with spherical confinement. To increase the sphericity of the formed cluster, several anti-Mackay shells[5,6,32,33] may be incorporated into the cluster structure.[13] Unlike atomic clusters whose surface atoms can be added or removed to reach a suitable minimum energy configuration by evaporation and condensation,[7] the number of colloidal particles in the confining emulsion droplet that templates colloidal clusters remains fixed. This invariance of particle numbers in colloidal clusters provides an ideal opportunity to study the complete number–structure–property relationship in finite confined self-assembling systems. While magic numbers are characterized by a minimum free energy with few structural defects, in most experiments the number of colloidal particles lies between two adjacent magic numbers. It is therefore of fundamental interest to understand how structure and free energy evolves as the number of colloidal particles transitions from one magic number to the next.

Here, we systematically investigate the free energy landscape of large colloidal clusters in the size range from ~1000 to ~8000 colloidal particles. Systems in this size range are known to reproducibly form clusters with icosahedral or near-icosahedral symmetry.[13,14] We investigate the number–structure–property relationship by analyzing the structure of colloidal clusters as a function of their system size, both in the vicinity of magic numbers as well as in-between regions of magic numbers. We use droplet-based microfluidics to confine submicron polystyrene (PS) particles inside emulsion droplets and vary their concentration to prepare colloidal clusters of different sizes. We complement these experiments with event-driven molecular dynamics (EDMD) simulations of hard spheres in rigid spherical confinement. In contrast to experiment where the exact number of particles can only be inferred, simulations allow the accurate study of size-dependent structure and thermodynamics. The systematic studies reveal that in the colloidal realm, the magic number minima in the free energy landscape are significantly broader than expected from atomic clusters. With the help of geometric modelling, we propose three mechanisms to account for this broadening. The mechanisms generate clusters with well-defined icosahedral concentric shells in the cluster core but a more variable structure in outer shells. Furthermore, in-between magic number regions, where the number of constituent building blocks does not allow the formation of a complete icosahedral structure, we observe a defect localization. Instead of being uniformly distributed throughout the cluster, defects accumulate into a wedge that covers only a few tetrahedral grains in a plane containing a near-five-fold axis. The accumulation of defects in the wedge breaks icosahedral symmetry by opening up one of the five-fold symmetries by a few degrees. The formation of this wedge reduces the free energy penalty associated with the presence of defects as the majority of the particles in the cluster remain defect-free and stress in tetrahedral grains is relaxed.

## RESULTS AND DISCUSSION

**Free energy landscape of colloidal clusters as a function of system size**

We produce colloidal clusters by slowly removing water from monodispersed emulsion droplets of a dispersion of 244 nm charge-stabilized PS particles. In the course of water removal, the volume fraction of colloidal particles increases, eventually triggering an ordering phase transition. During the last stage of drying, capillary forces consolidate the mobile colloidal particles into a solid cluster with icosahedral structure.[13] In computer simulations, we adopt a two-step scheme to mimic the drying process. First, event-driven molecular dynamics (EDMD) simulations of hard spheres in spherical confinement replicate the phase transition into the icosahedral structure with increasing volume fraction. Second, a Morse potential is applied to the system after the phase transition. The introduction of attractive interactions relaxes the structure and facets the cluster surface, mimicking the capillary forces consolidating the cluster in experiments.

Importantly, the complex structure of the self-assembled cluster already completely forms in the first step of the simulation scheme, while the second step only consolidates the cluster and causes the formation of flat planes at the icosahedral facets, which facilitates comparison to the experiment. The structure formation in the hard sphere simulation and the close agreement to the experimentally observed structures indicates that the assembly process is dominated by entropy. Interaction forces acting between the particles, such as the DLVO forces in our charge-stabilized colloidal dispersion, do not seem to be of importance for the structure formation in this confined self-assembly process.

Using EDMD simulations, we have previously shown that the phase transition results in a pressure drop of different magnitude when particle number is slightly varied.[13] The sensitive response of pressure to system size hints at a strong dependence of free energy on system size. Exemplary cluster structures are shown in Fig. 1a, where experimentally observed colloidal clusters in Scanning Electron Microscopy (SEM) are compared with clusters formed in simulation. The agreement between the two structures is high given the simple hard sphere model in the first step of the simulation and the rather coarse approximation of capillary forces by the Morse potential. Geometrically, magic colloidal clusters can be described by a Pentakis dodecahedron sphere packing model, which involves core Mackay shells and surface anti-Mackay shells (Figure S1).[13] We recently proposed the $(m+a)_a$ notation to describe the shell structure of magic colloidal clusters, where $m$ is the number of Mackay shells and $a$ is the number of anti-Mackay shells. The cluster in Fig. 1a (left) has 8 Mackay shells ($m$) and one anti-Mackay shells ($a$), hence the notation $9_1$. Fig. 1a (right) shows a cluster with 8 Mackay shells and 3 anti-Mackay shells, hence the notation $11_3$.

We obtain the free energy landscape of colloidal clusters, shown in Fig. 1b, by numerically calculating the free energy of hard sphere clusters in spherical confinement. For this purpose, we use configurations from EDMD simulations at 52% packing fraction and employ normalization to obtain the entropy per particle as described in our recent work.[13] In hard sphere systems, free energy is determined solely by entropy. Ideal Mackay and anti-Mackay clusters are constructed from our model by spherically truncating the Mackay icosahedron and Pentakis dodecahedron model (Figure S2). The free energy landscape of colloidal clusters shows distinct minima. These minima can be assigned to regions where the number of constituent particles in the colloidal cluster affords closed shells, and subsequent minima have increasing numbers of total shells. The free energy minima of simulated clusters coincide with the particle numbers required in the ideal model, suggesting that clusters with concentric icosahedral shells, as described by the model, are thermodynamically favored.

Noteworthily, the free energy curves of colloidal clusters from simulation do not show sharp peaks but rather broadened regions of low free energy (termed magic number regions, green in Fig. 1b), followed by plateaus with higher free energy (termed off-magic number regions, red in Fig. 1b). The appearance of broad magic number regions is surprising at first because varying the spherical truncation radius of the model to include more particles generates clusters with isolated particles at the cluster surface (Figure S2). Such particles lower the density and prevent efficient packing in spherical confinement, which increases the free energy. Apparently, the simple picture of sparsely located, discrete magic number common in the study of atomic clusters is not applicable to colloidal clusters.

We use bond orientational order diagrams (BODs) to assess the structural order in colloidal clusters. BODs provide information about the symmetry and arrangement of particles by projecting all bonds between neighboring particles to a spherical surface. In short, crystalline arrangements of particles generate BODs with clear and bright spots where the intensity of a spot depends on the number of aligned bonds. A detailed analysis of the BODs of model icosahedral clusters is provided in Figures S3-S7.

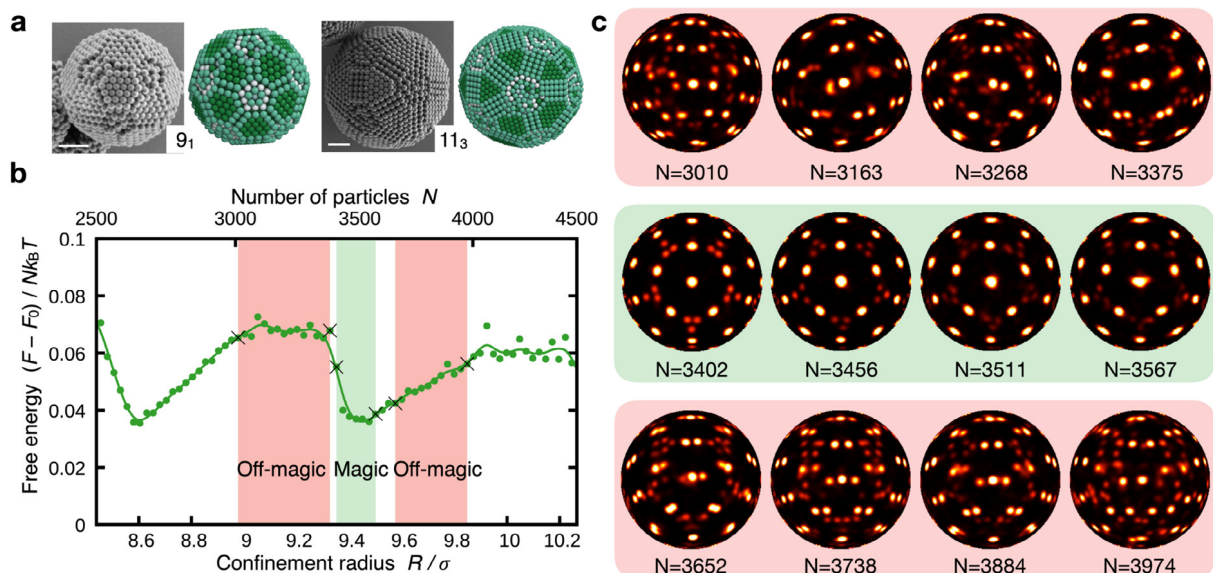

**Figure 1.** Magic and off-magic number regions in the free energy landscape of colloidal clusters. **a**, Two exemplary icosahedral magic number colloidal clusters from experiments (left) and simulation (right). The cluster surface is tiled with alternating triangles and rectangles around five-fold axes. The $9_1$ type cluster consists of nine icosahedral concentric closed shells among which there is one anti-Mackay shell at the surface. The $11_3$ type cluster consists of a total of eleven shells, including three anti-Mackay surface shells. Scale bars: 2 μm. **b**, Free energy landscape of clusters from between 2500 to 4500 colloidal particles at packing fraction 52% in simulations. Line is a guide to the eyes. As the number of particles in the cluster increases, two free energy minima are observed that correspond to nine and ten closed shells, respectively. Off-magic and magic number regions are marked in red and green, respectively. **c**, Bond orientational order diagrams (BODs) of selected simulated colloidal clusters. The free energy of the chosen clusters transitions from a plateau of high free energy through a minimum to another plateau. For the off-magic number region (3010 to 3375, red, top), individual spots in BODs split into duplets. For the magic number region (3402 to 3567, green, middle), BODs show clear and bright spots with five-fold patterns indicative of high-quality icosahedral ordering. When transitioning out of the free energy minimum into the next off-magic number region (3652 to 3974, red, bottom), BODs again show splitting of spots.

Fig. 1c shows BODs of several simulated colloidal clusters from about 3000 to 4000 particles in simulation following the transition in free energy between two high free energy plateaus through a minimum. Within the free energy minimum (particle numbers from 3402 to 3567), colloidal clusters exhibit near-perfect icosahedral symmetry (middle row in Fig. 1c). Their BODs show precise and clear spots with five-fold symmetric pattern. In addition, five triplets of subspots are found between the major spots with a decreasing intensity as cluster size increases. The structural origin of the subspots will be discussed below. Colloidal clusters outside of the free energy minimum (off-magic number) show broken icosahedral symmetry where spots in the BOD are split into doublets (particle numbers of 3010 to 3375 and 3652 to 3974; first and third row in Fig. 1c).

**Magic number regions**

We discuss three mechanisms that broaden minima in the free energy landscape and explain how icosahedral clusters can form throughout an extended region of magic numbers and do not exist as individual, discrete magic numbers only.

*Minimum broadening due to anti-Mackay shells.* The number of anti-Mackay shells sets upper and lower bounds for magic number regions. Fig. 2a shows a typical magic number colloidal cluster found in experiment together with the corresponding $10_2$ cluster model (Fig. 2b). The number of anti-Mackay shells can be identified from the structure of the cluster surface.[13] Five triangular patterns with close-packed configuration of particles in {111} planes surround the vertex regions and are separated by rectangular patterns with square configuration of particles in {110} planes. The appearance of rectangular patterns with three rows in Fig. 2a,b indicates that the cluster has two anti-Mackay shells.

Maintaining ten closed shells in the cluster, the number of anti-Mackay shells can vary (Fig. 2b-e).[13] Because rectangular regions in anti-Mackay shells accommodate fewer particles compared to Mackay shells, their presence decreases the packing density. Indeed, the number of particles decreases from 3607 in the cluster with no anti-Mackay shells ($10_0$ type cluster, Fig. 2c) to 3247 in the cluster with three anti-Mackay shells ($10_3$ type cluster, Fig. 2e). For well-formed icosahedral clusters, we always observe 42 bright spots in BODs in the directions of the vertices of a pentakis icosidodecahedron. These spots originate from 20 (slightly deformed) fcc grains in icosahedral arrangement (Fig. 2c, Fig. S3-S7). Anti-Mackay shells, which are twinned with Mackay shells, generate 20 triplets of subspots (Fig. 2b,d,e). Subspots become brighter with increasing number of anti-Mackay shells (Fig. 2b-e).

The intensity of the subspots in BODs of simulated clusters within the magic number region (second row in Fig. 1c) gradually diminishes with increasing particle number. This indicates a reduction of the number of anti-Mackay shells with cluster size. Different combinations of Mackay shells and anti-Mackay shells correspond to distinct magic numbers within a single magic number region. However, the free energy landscape does not resolve these sub-features within the free energy minimum, indicating additional mechanisms to accommodate variations in particle number.

*Minimum broadening due to partial anti-Mackay shell.* Additional types of ordered colloidal clusters are possible by adopting a partial anti-Mackay shell. Fig. 3a shows a simulated cluster in the magic number region. Its surface shows the characteristic rectangular patterns of the anti-Mackay shells. Noteworthily, rectangles with a width of both three and four particles can be seen. The joint appearance of these two rectangular patterns suggests that grains with two and three anti-Mackay shells coexist. Indeed, the BOD shows asymmetric subspot triplets with uneven intensity, corroborating an anisotropic thickness of anti-Mackay shells at the cluster surface. Fig. 3b shows a side-view of a grain extracted from the cluster shown in Fig. 3a with seven Mackay shells in the core and three anti-Mackay shells at the surface. Note that our notation considers the central sphere in the cluster as the zeroth Mackay shell. Fig. 3c shows another grain from the same cluster with eight Mackay shells and two anti-Mackay shells. Partial anti-Mackay shells are also frequently observed in experiment (Fig. 3d,e).

Recall that for a fixed total number of shells, grains with a higher number of anti-Mackay shells consist of fewer particles (Fig. 2). Given there are now two types of grains that coexist within a cluster (Fig. 3b,c), it becomes clear that partial anti-Mackay shells increase the types of ordered clusters that comprise the magic number region.[9] Note that in a strict sense icosahedral symmetry is disrupted by adapting a partial anti-Mackay shell. However, because the underlying core structure remains unaffected, such clusters have similar free energies as magic number clusters with complete icosahedral symmetry. The formation of partial anti-Mackay shells smears out the discrete free energy minima arising from Mackay and anti-Mackay clusters.

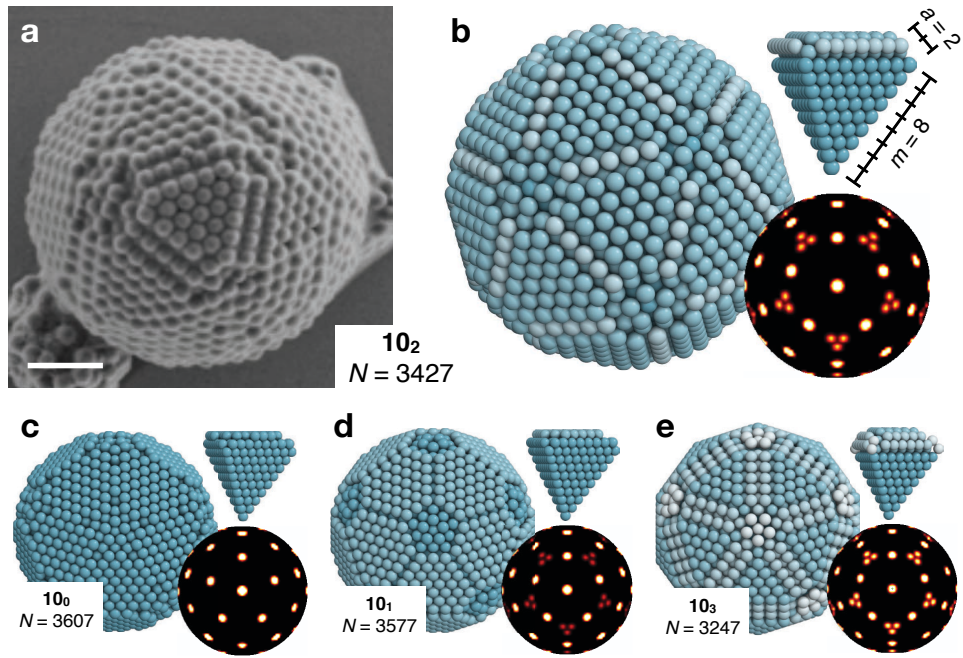

**Figure 2.** Ten-shell colloidal clusters with variable number of anti-Mackay shells. **a,b**, Scanning electron microscope (SEM) image of experimentally observed colloidal cluster (a) and corresponding model (b). Scale bar: 2 μm. The cluster is identified as a $10_2$ type with 3427 colloidal particles. One of the 20 tetrahedral grains and the BOD of the cluster are shown as inset. **c**, The BOD of the 10_0 cluster consists of 42 bright spots along the vertices of a pentakis icosidodecahedron. **d,e**, BODs of the 10_1 cluster (**d**) and the 10_3 cluster (**e**) have additional triplets of less intense subspots. These subspots result from the anti-Mackay surface shells. With increasing number of anti-Mackay shells from (**c**) *via* (**d**) and (**b**) to (**e**) the intensity of subspots increases proportional to the total number of particles in all anti-Mackay shells.

*Minimum broadening due to disorder near cluster vertices*. The areas near the vertices pointing along five-fold symmetry axes are usually more disordered than the otherwise regularly tiled surface of magic number clusters. Fig. 4a shows a typical colloidal cluster with a surface exhibiting rectangular patterns with a width of four particles and a side-length of seven particles (marked in blue). The corresponding model structure, composed of 9063 spheres, identifies the cluster as $14_3$ type (Fig. 4b). Fig. 4c shows a different colloidal cluster in the experiment along with its model (Fig. 4d). As evidenced from the width of the surface rectangular patterns of four particles, this cluster also consists of three anti-Mackay layer and therefore is equally classified as $14_3$ type. However, two lines of these rectangular patterns are removed at both sides leading to a total side-length of five particles. The similarity of the two clusters is obvious from their BODs, which are indistinguishable (Fig. 4b,d, insets). Subtle differences in the vertex region are not captured by our cluster classification. In the model, both clusters can be generated by different truncation radii (Figure S2f,h).

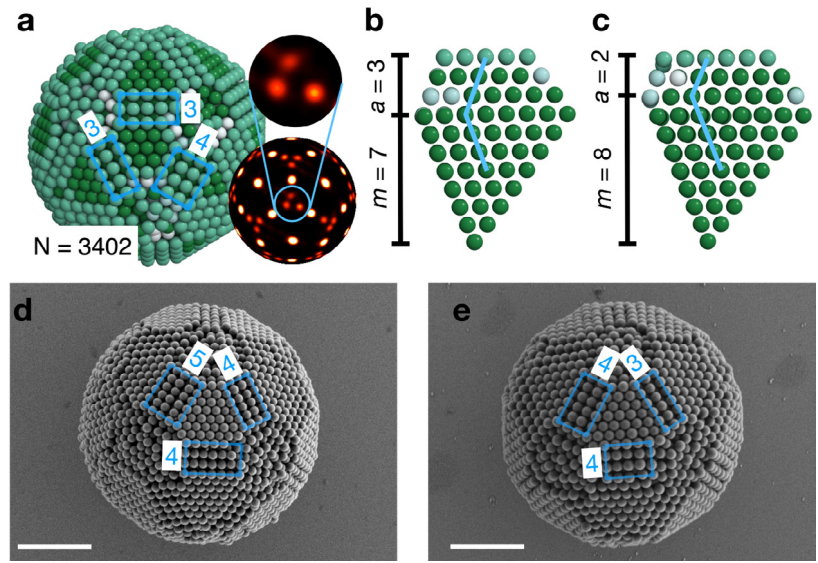

**Figure 3.** Colloidal clusters with partial anti-Mackay shell. **a**, Simulated cluster within the magic number region (3402 spheres) that exhibits a surface structure with rectangular patterns of a width of both three and four particles, indicating the presence of different numbers of anti-Mackay shells within the same cluster. The uneven intensity of triplet subspots in the BOD (inset) results from an anisotropic thickness of anti-Mackay shells. **b,c**, Side-view of two grains extracted from (**a**). Anti-Mackay twinning is indicated by blue lines. Spheres are drawn slightly smaller for clarity. **d,e**, Experimental observation of two colloidal clusters with a partial anti-Mackay shell. Scale bar: 2 μm.

The vicinity of the vertices are typically the most deformed parts of icosahedral clusters.[13] Defects near the vertices therefore have the lowest free energy cost. In both experiments and simulations, these regions typically exhibit several extra or missing particles, while the overall structure remains unaffected. In the example of Fig. 4, such excess particles near the vertices amount to 360 particles, about 4% of the total number of particles.

All three mechanisms jointly contribute to the broadening of minima in the free energy landscape of colloidal clusters. Whenever possible, self-assembly in spherical confinement attempts to maintain icosahedral symmetry with closed, concentric shell structure. System size fluctuations up to several hundred particles can be accommodated efficiently *via* changes in the number of anti-Mackay layers, the introduction of partial anti-Mackay layers, and excess particles in the vicinity of the cluster vertices. However, these mechanisms are not sufficient to explain the plateau regions of higher free energy in the off-Mackay regions. BODs of colloidal clusters in these regions exhibit a broken icosahedral symmetry, often with smeared out spots suggesting the presence of defects within the core region (Fig. 1c). We now address the question how these defects are incorporated and distributed within the cluster to minimize their detrimental effects to free energy.

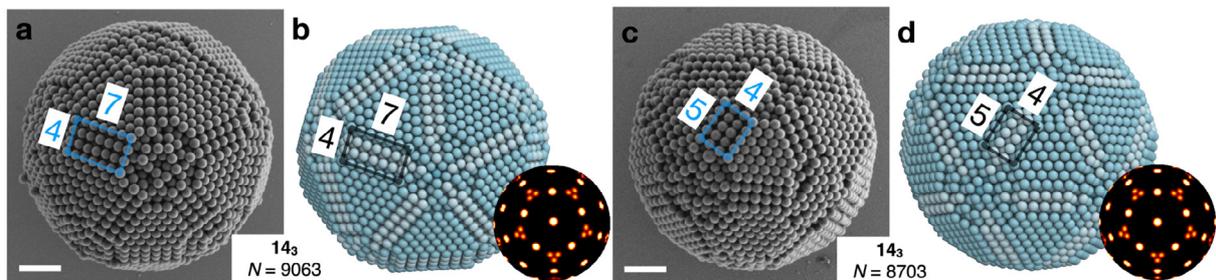

**Figure 4.** Excess or deficient particles near the cluster vertices. **a,c**, Magic number colloidal clusters with two different side-length of rectangular surface patterns but equal width. Both clusters can be assigned a $14_3$ type. Scale bars: 1 μm. **b,d**, Corresponding geometric models and BODs (inset). The model and the indistinguishable BODs suggest that the two clusters only differ near the vertices but are identical in the core region. Both BODs are rotated to their five-fold axes for clarity.

**Off magic number regions**

Fig. 5a shows a magic number cluster obtained from simulation using a number of particles near the center of a free energy minimum ($N = 3402$ in Fig. 1). The cluster is viewed along three different symmetry axes. The cluster surface appears well ordered, and the BODs exhibit clear and bright spots. All characteristic surface features predicted by the anti-Mackay model are present. In contrast, Fig. 5b shows an off-magic number cluster with only a few hundred more particles such that the number of particles is now outside of the broadened free energy minimum ($N = 3738$ in Fig. 1). Again, the cluster surface appears well ordered in most orientations without clear visual distinction from the magic number cluster above. Still, close inspection reveals signs of disorder when viewed along one of the two-fold axes (Fig. 5b, middle). Differences are much more pronounced in the BODs. Clear splitting of BOD spots, typically into doublets, indicate breaking of icosahedral symmetry in the off-magic number cluster. Apparently, the relative orientation of tetrahedral grains (Fig. S1, S3-S7) is disturbed. From this example it is obvious that crystalline order of colloidal clusters should not solely be evaluated from analysis of the cluster surface.

We investigate the distribution of defects in the simulated clusters. After removing surface particles and particles in anti-Mackay shells that have reduced numbers of neighbors, we identify particles with less than 12 nearest neighbors as defects. While the magic number cluster of Fig. 5a contains no defects in this analysis (not shown), the off-magic number cluster accumulates defects in a narrow wedge (Fig. 5b, red particles in inset). We find that this wedge takes up the volume of between three and four tetrahedron grains. The accumulation of defects in the wedge leaves non-affected grains in the cluster defect-free.

We model the orientation and contact of tetrahedron grains in a magic number cluster and an off-magic number cluster. The magic number (Mackay) cluster has 20 large grains with icosahedral symmetry (Fig. 5c). The distance between icosahedron vertices in the model is 1.05 times larger than the distance from the center to the vertices. Internal stress in the Mackay cluster is inevitable because tetrahedron grains with dihedral angle $acos\,(1/3) = 70.53°$ do not quite fit pentagonal symmetry with rotation angle $360\,°/5 = 72°$. Magic number clusters accept the free energy penalty associated with this stress because it is more than compensated by the highly efficient surface packing that adopts well to spherical confinement.

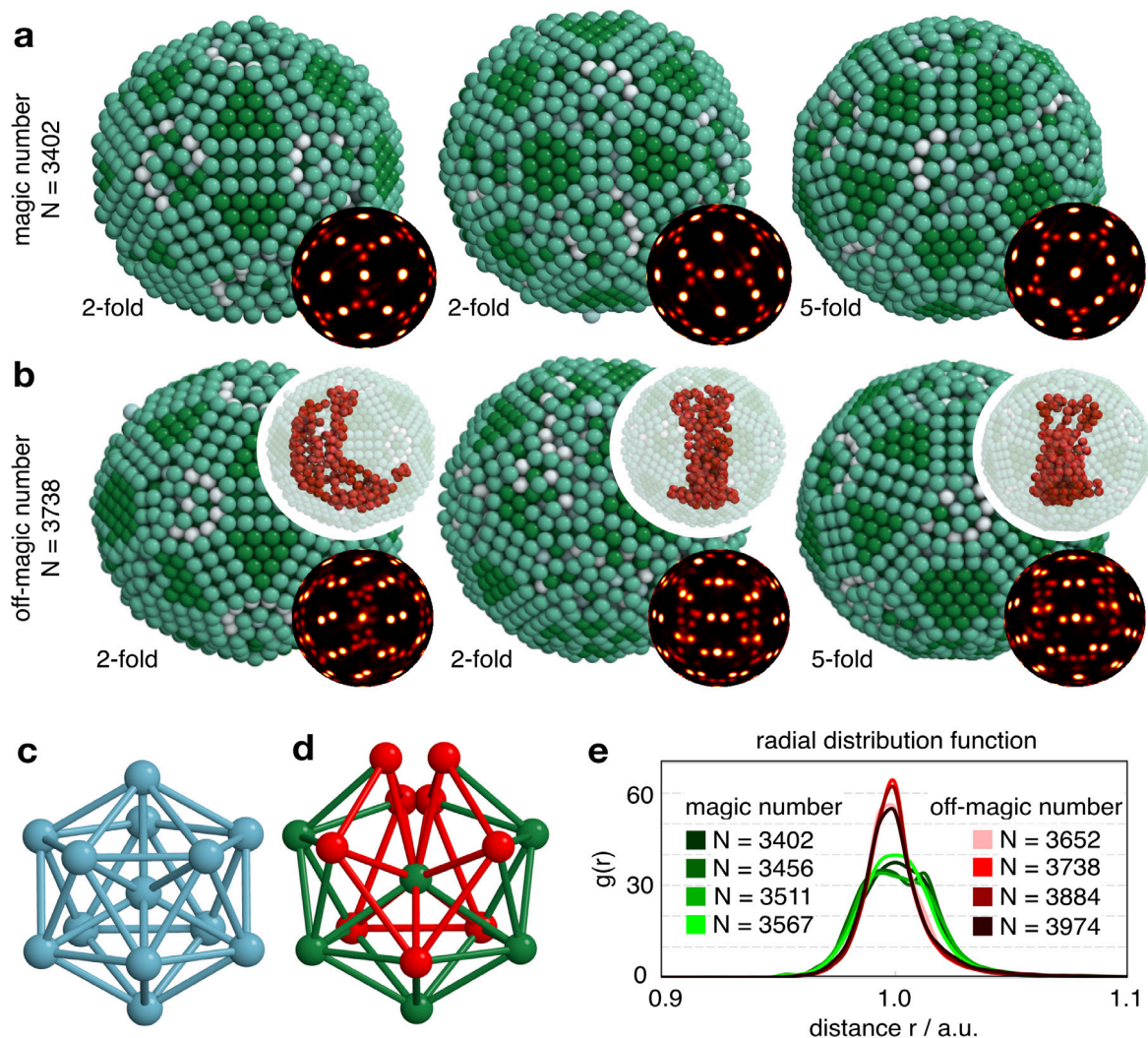

**Figure 5.** Defects accumulate in a wedge for simulated off-magic number cluster. **a**, An exemplary simulated magic number cluster (3402 spheres) viewed along two two-fold and one five-fold axis. BODs (insets) show sharp, bright spots with near-perfect icosahedral symmetry. **b**, A simulated off-magic number cluster (3738 spheres) has similar order on the surface but BODs now show duplets of split spots indicating broken icosahedral symmetry. Top right insets indicate the distribution of defect particles (marked red) inside the cluster (semi-transparent). Defects accumulate in a wedge perpendicular to a two-fold axis (left), parallel to a two-fold axis (middle) and parallel the five-fold axis (right). Defect particles near the surface and the different geometry of the anti-Mackay shells were ignored by analyzing only particles located within 75% of the cluster radius. **c**, Perfect icosahedron model for the Mackay (and anti-Mackay) cluster. **d**, Defected icosahedron model obtained by minimizing the stress between tetrahedron grains. An inevitable gap (approximate size three tetrahedra, boundary marked in red) accommodates the accumulated defects. **e**, First peak of the radial distribution function of magic number clusters (green colors) and off-magic number clusters (red colors).

The situation is different for off-magic number clusters (Fig. 5d). We construct a defected icosahedron model starting from a pentagonal bipyramid whose 15 edges are of equal length. Regular tetrahedra are subsequently added until overlap becomes inevitable (Fig. S8) and a large opening (marked in red) is left. The opening covers a volume of approximately three more tetrahedra. The opening, which accommodates the defects, has a wedge-like shape in which two twinned tetrahedra connect to a third one through a narrow gap. Movie S1 and an interactive Javascript visualization, available as Supplementary Information, show the three-dimensional rotation of the defected icosahedron model.

The defected icosahedron model accurately reproduces the split spots in BODs of off-magic number clusters (Fig. S9). The defect wedge forces the cluster into two halves, each slightly tilted from its original position in the Mackay cluster. According to the defect model and in agreement with observations in simulated clusters, there remains no five-fold symmetry axis in the structure but only mirror planes. The sacrifice of icosahedral symmetry in the defect model relaxes internal stress by leaving the remaining grains less deformed. Relaxation of stress is directly visible in the radial distribution function (Fig. 5e). In this plot, the first peak, corresponding to nearest neighbor distances, is much narrower for off-magic number clusters (red) than for magic number clusters (green). Although our calculations suggest that the transition from magic number region to off-magic number region is gradual (Fig. 1b), the amount of peak splitting is always very similar. This suggests that the opening angle of the wedge in the defected icosahedron model is unique and likely determined by geometry.

We confirm the existence of accumulated defects in experiment by means of 360° electron tomography (ET) in scanning transmission electron microscopy (STEM) mode. After deposition of colloidal clusters on a Lacey carbon grid, we select a cluster with highly-ordered surface pattern and transfer it onto a tip with a plateau matching the respective particle size. The freestanding particle-on-tip geometry enables 360° ET, which allows for high-precision 3D analyses without "missing wedge" artefacts.[34] Details about the transfer procedure can be found in reference.[35] We obtain 180 STEM images of the selected sample with 1° tilt increments. Each image represents a single projection from the sample's interior 3D structure onto a 2D plane along the respective imaging direction. The 3D reconstruction of the entire tilt series shows the full 3D mass distribution of the selected colloidal cluster (Movie S2). The surface rendering of the reconstructed colloidal cluster (Fig. 6a) shows clear local five-fold patterns and anti-Mackay features with rectangular patterns of width three particles (Movie S3). The colloidal cluster is viewed along a two-fold axis and contains about 5000 particles, as determined manually in Visual Reality (VR) visualization software.

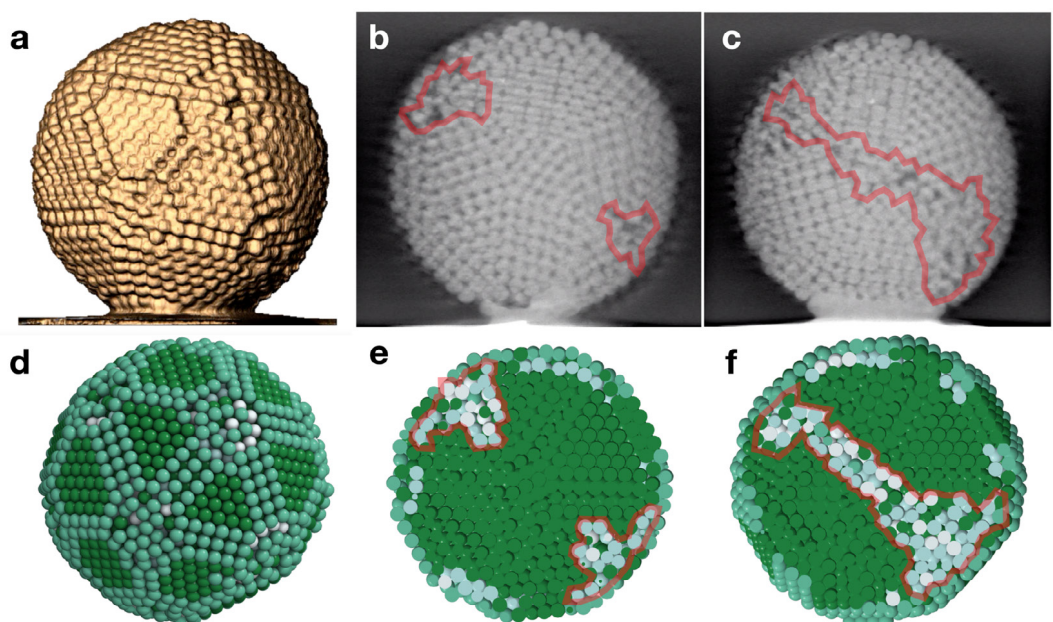

**Figure 6.** Electron tomographic reconstruction of a colloidal cluster confirms the accumulation of defects in a wedge. **a**, Surface rendering of electron tomographic reconstruction of colloidal cluster with a primary particle size of about 190 nm. The typical surface tiling with rectangular patterns (width three particles) characteristic for highly ordered colloidal clusters is observed. **b**, Virtual slice through the reconstructed cluster at its upper hemisphere perpendicular to a two-fold axis reveals the interior structure with eight ordered grains. Two small disordered regions (marked by red lines) are visible at opposing sides. **c**, Slice at the lower hemisphere. Only six ordered grains are seen, covering a reduced area. The defect region (marked by red lines) extends from two opposing sides (present in (**b**)) towards the center separating the ordered regions into two parts. The defect region lies along the two-fold axis (oriented in viewing direction). **d**, A simulated off-magic number cluster (5088 spheres) is shown for comparison. Particles are colored by the number of nearest neighbors. Particles in dark green are crystalline with 12 neighbors. Particles in white and light green are in defect regions with lower numbers of neighbors. **e**,**f**, Slices through the upper (**e**) and lower (**f**) hemisphere reveal a wedge-like defect region, agreeing with the observations in the reconstructed experimental colloidal cluster (**b**,**c**).

The interior structure of the cluster is revealed by slices at different positions through the three-dimensional reconstruction. The slice at the upper hemisphere cuts through eight domains of fcc grains (Fig. 6b). Most particles are in crystalline domains including two grains showing {111} crystal planes. The revealed plane exhibit overall two-fold symmetry. Two small disordered regions at opposing sites are marked by red lines in Fig. 6b. As the slicing plane moves towards the lower hemisphere, the disordered region extends towards the middle (Movie S4). Once the slice reaches the lower hemisphere (Fig. 6c), the defect regions (marked in red) join through the center of the image that separates the crystalline domains, corroborating with the wedge-like distribution of defects predicted from the model.

A simulated cluster of 5088 particles, shown in Fig. 6d, has identical structural features as the experimentally investigated cluster underlining that our two-step simulation scheme accurately replicates the experiment. Cross-sections of this simulated cluster taken at identical heights as the tomographic data reveal a similar distribution of defect (Fig. 6e,f, Movie S5). The upper hemisphere of the simulated cluster shows a high degree of crystallinity (Fig. 6e). Most particles are in crystalline domains (marked dark green). In the lower hemisphere, the defect wedge appears across the cluster (Fig. 6f). Overall, we conclude that the position of the defect wedge agrees well between experiment, simulation and geometric model. Additional analyses of interior structures of different clusters in magic number and off-magic number regions can be found in Movie S6 and Fig. S10-S13.

We systematically analyze all simulated clusters containing between 1000 and 5000 particles. For off-magic number clusters defects accumulate in a wedge as described above whereas in magic number clusters much fewer defects are present. We plot the fraction of defect particles as a function of system size and compare it to the free energy landscape in Fig. 7. The fraction of defects fluctuates as cluster size increases and is minimal in the magic number regions where free energy has a minimum. As the system size transitions out of a magic number region, the fraction of defect particles drastically increases. The correlation between the fraction of defects in the hard-sphere simulations and the free energy landscape underlines the entropy-driven magic number effect in colloidal clusters.

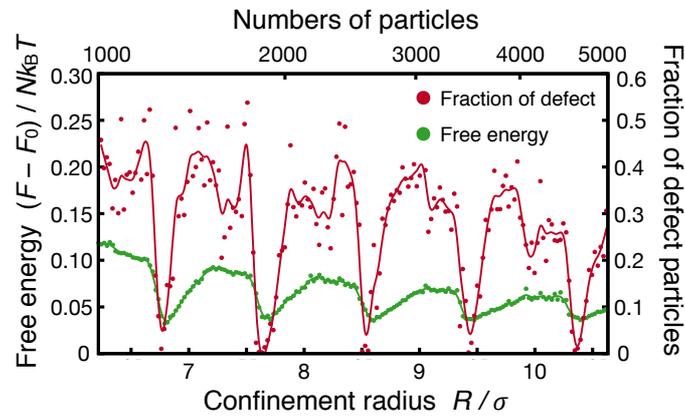

**Figure 7.** The fraction of defect particles in colloidal clusters is correlated to the free energy landscape. The calculated free energy of colloidal clusters is shown in green, the fraction of defect particles is shown in green. Only particles in the cluster interior (75% of the radius) are used for the analysis to exclude effects of surface tiling. Interpolated lines are guides to the eye. In magic number regions, the fraction of defect particles is low. In off-magic number regions, corresponding to higher free energy, the clusters contain a higher fraction of defect particles.

## CONCLUSION

Confined self-assembly in a spherical droplet produces colloidal clusters with icosahedral symmetry and concentric closed shell structure as minimum free energy structures. The close agreement of symmetry and details of the internal structure between experiments using charge-stabilized colloidal particles and hard sphere simulations indicate that entropy is the dominant force governing the assembly process. Geometric rules only allow certain numbers of spheres to assembly perfectly icosahedral concentric shell structures, just as 14 spheres cannot pack into an icosahedron. We analyzed the free energy landscape of colloidal clusters as a function of system size and identified regions with low free energy that correspond to magic number clusters with closed shell structure. In contrast to atomic magic number clusters that show discrete free energy minima, the free energy minima in colloidal clusters are broadened. We proposed three mechanisms that contribute to the broadening of minima in the free energy landscape of colloidal clusters. Whenever possible, our confined self-assembling system attempts to maintain global icosahedral symmetry with closed, concentric shells but can efficiently accommodate fluctuations up to several hundred particles *via* adjustments within anti-Mackay layers, mixing anti-Mackay layers, and fluctuations in order near the vertices of the cluster.

Outside the magic number regions, icosahedral symmetry is clearly broken. In these regions, the system accumulates defects in a wedge separating grains instead of evenly distributing stress throughout all 20 grains. We proposed a defected icosahedron model to account for the distribution of defects in off-magic number clusters. The tomographic reconstruction of an experimentally obtained cluster and defect analysis of simulated clusters confirm this defect accumulation. The fluctuation of free energy is shown to be associated with the fraction of defect particles in the cluster.

The structures of finite self-assembling systems is a complex function of shape,[36–41] composition[42–45] and interaction[46–51] of the particles. We demonstrated the use of magic numbers as a tool to understand structures in finite confined systems, even at a seemingly large system size.

## METHODS

**PS nanoparticles and colloidal cluster synthesis**. Polystyrene (PS) nanoparticles were synthesized by using ammonium peroxodisulfate as initiator and acrylic acid as comonomer in surfactant-free emulsion polymerization.[52] Colloidal clusters were fabricated by drying of monodispersed water in oil emulsion droplet produced in PDMS microfluidic device. The microfluidic device was produced by soft photolithography as described in literature.[18] In short, the silicon wafer was patterned by exposing thin layer of negative photoresist SU-8 by UV light through designed photomask. Polydimethylsiloxane (Sylgard 184, Dow Corning) was mixed 10:1 ratio with curing agent and poured onto the silicon master to replicate the microstructure. The PDMS was cured in 85°C overnight and later peeled off, punched with 1mm diameter biopsy punch for inlets and outlets. The PDMS chip was bonded to glass slide after 30 W oxygen plasma for 18 s. The channels were flooded by Aquapel (PPG Industries) to induce hydrophobicity. An aqueous dispersion of polystyrene particles (d = 244nm) was pumped into the microfluidic chip to produce emulsion droplets. 0.1 wt % PFPE-PEG-PFPE surfactant, synthesized following literature,[53] was dissolved in perfluorinated carbon oil (3M Novec Engineering Fluid HFE 7500) to stabilize the emulsions. Emulsion droplets were collected in 1.5 ml glass vial to store and dried.

**Electron microscopy and tomography.** Colloidal clusters were drop-casted onto a silicon wafer. SEM images were obtained by Zeiss Gemini Ultra 50 at 1 kV electron voltage. For applying 360° electron tomography (ET) colloidal clusters were drop-casted from solution onto a standard (200 mesh) Lacey carbon copper grid. The grid was mounted in the SEM (FEI Helios NanoLab 660) to identify a suitable structure and cluster size for the three-dimensional (3D) analysis. A colloidal cluster with a size of about 3.3 $\mu$m containing about 5000 primary particles was selected and transferred onto a tailored tomography tip by the stamping transfer technique, as described in reference,[35] using SEM-compatible glue do increase the adhesion between tip and particle. The tomography tip with the attached particle was mounted onto a Fischione Model 2050 On-Axis Rotation Tomography holder (E.A. Fischione Instruments, Inc.) and transferred to the transmission electron microscope (TEM). 360° Scanning transmission electron microscopy (STEM) tomography was performed using a dual probe- and image-side aberration-corrected FEI Titan3 Themis 60-300 transmission electron microscope at an acceleration voltage of 300 kV in high-angle annular dark field (HAADF) STEM imaging mode at a camera length of 91 mm. The semi-convergence angle of the STEM probe (microprobe STEM) was lowered to 0.44 mrad to increase the depth of field

(DOF) to image all parts of the sample completely in focus throughout the entire tilt series acquisition procedure.[54] The diffraction-limited resolution for the adapted semi-convergence angle was 2.7 nm at a respective DOF of 7.6 $\mu$m. The large sample size caused a strong decrease of resolution due to broadening of the STEM probe (multiple elastic scattering),[55] which leads to an estimated STEM probe diameter of about 90-120 nm. The tilt series was acquired using FEI Tomography 4.0 software within a full tilt angle range of 180° with 1° tilt increment, continuous and linear tilting scheme and auto focus and tracking before acquisition (Movie S2). To prevent morphological changes, *e.g.* shrinkage, of the sample during tilt series acquisition a low beam current of 60 pA was applied and the sample was illuminated for 10 minutes before performing the measurement. The tilt series was aligned using FEI Inspect 3D software (cross-correlation technique). The tomogram was reconstructed with the simultaneous iterative reconstruction technique (SIRT)[56] applying 50 iterations using FEI Inspect 3D software. For 3D data analyses the reconstructed volume was visualized with VSG Avizo 8.1 for FEI systems software. A median filter was used to minimize background noise, and a global threshold value was applied to segment the particles from the pore space. Numbers of particles were counted manually in visual reality (VR) software in Arivis InViewR.

**Simulation of colloidal clusters and free energy calculation**. Our previously developed two-step simulation procedure[13] was utilized to replicate colloidal clusters observed in the experiments. In the first step, colloidal particles were modelled as hard spheres with diameter $\sigma$ in hard spherical confinement of radius $R$. Evaporation was simulated with event-driven molecular dynamics in the isochoric (*NVT*) ensemble by gradually increasing packing fraction. In the second step, to mimic capillary forces that consolidate colloidal clusters during the final stage of droplet drying, hard sphere interaction was replaced with the Morse potential $V(r) = D_0 \left( e^{-2\alpha(r-r_0)} - 2e^{-\alpha(r-r_0)} \right)$, wherein $r_0 = \sigma$, $D_0 = 1$, $\alpha = 10$, and energy was minimized with the fast inertia relaxation engine (FIRE) implemented in HOOMD-blue.[57,58] Absolute free energies were calculated following our previous method[13] without modification based on the Frenkel-Ladd method[59] with particle swap moves[60] in a Monte Carlo simulation. For the reference system an Einstein crystal with harmonic springs $V(r) = \lambda(r - r_0)^2$ was used where $\lambda$ is the spring constant and $(r - r_0)$ the displacement from the spring anchor points $r_0$. The spring constant $\lambda$ was increased logarithmically in discrete steps over the range $10^{-5}$ to $10^5$. Absolute free energy values were obtained after subtracting the bulk contribution $F_0(N, \phi)$ to the free energy.[13] Particles in the simulated clusters were assigned colors based on their coordination number. Spheres in the interior region with 12 neighbors and spheres at the surface with 9 neighbors were assigned dark green. Spheres with coordination numbers between 4 and 8 were assigned medium green. Spheres with coordination number 11 were assigned light green, and the remaining spheres were assigned white.

## ASSOCIATED CONTENT

**Supporting Information**

Supporting information available. The following material is available free of charge *via* the Internet at http://pubs.acs.org.

Supporting Information

SI movie S1, Defect accumulation in simulated cluster, icosahedron and defect model

SI movie S2, Tilt series of STEM images of a colloidal cluster

SI movie S3, Surface view of the reconstructed colloidal cluster

SI movie S4, Slices through the reconstructed colloidal cluster

SI movie S5, Slices through a simulated off-magic number colloidal cluster

SI movie S6, Slices through simulated colloidal cluster with magic and off-magic numbers

## AUTHOR INFORMATION

**Corresponding Authors**

Michael Engel: michael.engel@fau.de, Nicolas Vogel: nicolas.vogel@fau.de

**Author Contributions**

The manuscript was written through contributions of all authors. All authors have given approval to the final version of the manuscript.

## ACKNOWLEDGMENTS

This work was supported by Deutsche Forschungsgemeinschaft (DFG) through the projects EN 905/2-1 and VO 1824/7-1 to M.E. and N.V., respectively. T.P., S.E. and E.S. acknowledge support by DFG within the framework of the research training group GRK 1896. All authors acknowledge the Cluster of Excellence Engineering of Advanced Materials Grant EXC 315/2, the Central Institute for Scientific Computing (ZISC), the Interdisciplinary

Center for Functional Particle Systems (FPS), and the Center for Nanoanalysis and Electron Microscopy (CENEM) at Friedrich-Alexander University Erlangen-Nürnberg. Computational resources and support provided by the Erlangen Regional Computing Center (RRZE) are gratefully acknowledged. B. Apeleo Zubiri is acknowledged for his advice regarding electron tomography and data analysis.

REFERENCES

(1) Echt, O.; Sattler, K.; Recknagel, E. Magic Numbers for Sphere Packings: Experimental Verification in Free Xenon Clusters. *Phys. Rev. Lett.* **1981**, *47*, 1121–1124.
(2) Doye, J. P. K.; Wales, D. J. The Effect of the Range of the Potential on the Structure and Stability of Simple Liquids: From Clusters to Bulk, from Sodium to C-60. *J. Phys. B- Atomic Mol. Opt. Phys.* **1996**, *29*, 4859–4894.
(3) Mottet, C.; Tréglia, G.; Legrand, B. New Magic Numbers in Metallic Clusters: An Unexpected Metal Dependence. *Surf. Sci.* **1997**, *383*, L719–L727.
(4) Baletto, F.; Ferrando, R. Structural Properties of Nanoclusters: Energetic, Thermodynamic, and Kinetic Effects. *Rev. Mod. Phys.* **2005**, *77*, 371–423.
(5) Mackay, A. L. A Dense Non-Crystallographic Packing of Equal Spheres. *Acta Crystallogr.* **1962**, *15*, 916–918.
(6) Kuo, K. H. Mackay, Anti-Mackay, Double-Mackay, Pseudo-Mackay, and Related Icosahedral Shell Clusters. *Struct. Chem.* **2002**, *13*, 221–230.
(7) Martin, T. P.; Bergmann, T.; Göhlich, H.; Lange, T. Observation of Electronic Shells and Shells of Atoms in Large Na Clusters. *Chem. Phys. Lett.* **1990**, *172*, 209–213.
(8) Martin, T. P.; Näher, U.; Bergmann, T.; Göhlich, H.; Lange, T. Observation of Icosahedral Shells and Subshells in Calcium Clusters. *Chem. Phys. Lett.* **1991**, *183*, 119–124.
(9) Hendy, S. C.; Doye, J. P. K. Surface-Reconstructed Icosahedral Structures for Lead Clusters. *Phys. Rev. B Condens. matter* **2002**, *66*, 235402.
(10) Martin, T. P. Shells of Atoms. *Phys. Rep.* **1996**, *273*, 199–241.
(11) Doye, J. P. K.; Calvo, F. Entropic Effects on the Size Dependence of Cluster Structure. *Phys. Rev. Lett.* **2001**, *86*, 3570–3573.
(12) Rahm, J. M.; Erhart, P. Beyond Magic Numbers: Atomic Scale Equilibrium Nanoparticle Shapes for Any Size. *Nano Lett.* **2017**, *17*, 5775–5781.
(13) Wang, J.; Mbah, C. F.; Przybilla, T.; Apeleo Zubiri, B.; Spiecker, E.; Engel, M.; Vogel, N. Magic Number Colloidal Clusters as Minimum Free Energy Structures. *Nat. Commun.* **2018**, *9*, 5259.
(14) de Nijs, B.; Dussi, S.; Smallenburg, F.; Meeldijk, J. D.; Groenendijk, D. J.; Filion, L.; Imhof, A.; van Blaaderen, A.; Dijkstra, M. Entropy-Driven Formation of Large Icosahedral Colloidal Clusters by Spherical Confinement. *Nat. Mater.* **2014**, *14*, 56–60.
(15) Peng, B.; Smallenburg, F.; Imhof, A.; Dijkstra, M.; Van Blaaderen, A. Colloidal Clusters by Using Emulsions and Dumbbell-Shaped Particles: Experiments and Simulations. *Angew. Chemie - Int. Ed.* **2013**, *52*, 6709–6712.
(16) Lauga, E.; Brenner, M. P. Evaporation-Driven Assembly of Colloidal Particles. *Phys.*

*Rev. Lett.* **2004**, *93*, 1–4.

(17) Agthe, M.; Plivelic, T. S.; Labrador, A.; Bergström, L.; Salazar-Alvarez, G. Following in Real Time the Two-Step Assembly of Nanoparticles into Mesocrystals in Levitating Drops. *Nano Lett.* **2016**, *16*, 6838–6843.

(18) Vogel, N.; Utech, S.; England, G. T.; Shirman, T.; Phillips, K. R.; Koay, N.; Burgess, I. B.; Kolle, M.; Weitz, D. A.; Aizenberg, J. Color from Hierarchy: Diverse Optical Properties of Micron-Sized Spherical Colloidal Assemblies. *Proc. Natl. Acad. Sci.* **2015**, *112*, 10845–10850.

(19) Velev, O. D.; Furusawa, K.; Nagayama, K. Assembly of Latex Particles by Using Emulsion Droplets as Templates. 2. Ball-Like and Composite Aggregates. *Langmuir* **1996**, *12*, 2385–2391.

(20) Velev, O. D.; Lenhoff, A. M.; Kaler, E. W. A Class of Microstructured Particles Through Colloidal Crystallization. *Science* **2000**, *287*, 2240–2243.

(21) Manoharan, V. N.; Elsesser, M. T.; Pine, D. J. Dense Packing and Symmetry in Small Clusters of Microspheres. *Science* **2003**, *301*, 483–487.

(22) Yi, G. R.; Manoharan, V. N.; Klein, S.; Brzezinska, K. R.; Pine, D. J.; Lange, F. F.; Yang, S. M. Monodisperse Micrometer-Scale Spherical Assemblies of Polymer Particles. *Adv. Mater.* **2002**, *14*, 1137–1140.

(23) Kim, S. H.; Lee, S. Y.; Yi, G. R.; Pine, D. J.; Yang, S. M. Microwave-Assisted Self-Organization of Colloidal Particles in Confining Aqueous Droplets. *J. Am. Chem. Soc.* **2006**, *128*, 10897–10904.

(24) Brugarolas, T.; Tu, F. Q.; Lee, D. Directed Assembly of Particles Using Microfluidic Droplets and Bubbles. *Soft Matter* **2013**, *9*, 9046–9058.

(25) Zhao, Y.; Shang, L.; Cheng, Y.; Gu, Z. Spherical Colloidal Photonic Crystals. *Acc. Chem. Res.* **2014**, *47*, 3632–3642.

(26) Lacava, J.; Born, P.; Kraus, T. Nanoparticle Clusters with Lennard-Jones Geometries. *Nano Lett.* **2012**, *12*, 3279–3282.

(27) Manoharan, V. N. Colloidal Matter: Packing, Geometry, and Entropy. *Science* **2015**, *349*, 1253751.

(28) Pusey, P. N.; van Megen, W. Phase Behaviour of Concentrated Suspensions of Nearly Hard Colloidal Spheres. *Nature* **1986**, *320*, 340–342.

(29) Frenkel, D. Order through Entropy. *Nat. Mater.* **2014**, *14*, 9–12.

(30) van Anders, G.; Klotsa, D.; Ahmed, N. K.; Engel, M.; Glotzer, S. C. Understanding Shape Entropy through Local Dense Packing. *Proc. Natl. Acad. Sci.* **2014**, *111*, E4812–E4821.

(31) Guerra, R. E.; Kelleher, C. P.; Hollingsworth, A. D.; Chaikin, P. M. Freezing on a Sphere. *Nature* **2018**, *554*, 346–350.

(32) Haberland, H.; Hippler, T.; Donges, J.; Kostko, O.; Schmidt, M.; Von Issendorff, B. Melting of Sodium Clusters: Where Do the Magic Numbers Come From? *Phys. Rev. Lett.* **2005**, *94*, 1–4.

(33) Doye, J. P. K.; Wales, D. J. Thermally-Induced Surface Reconstructions of Mackay Icosahedra. *Zeitschrift für Phys. D Atoms, Mol. Clust.* **1997**, *40*, 466–468.

(34) Midgley, P. A.; Weyland, M. 3D Electron Microscopy in the Physical Sciences: The Development of Z-Contrast and EFTEM Tomography. *Ultramicroscopy* **2003**, *96*, 413–431.

(35) Przybilla, T.; Zubiri, B. A.; Beltrán, A. M.; Butz, B.; Machoke, A. G. F.; Inayat, A.; Distaso, M.; Peukert, W.; Schwieger, W.; Spiecker, E. Transfer of Individual Micro- and Nanoparticles for High-Precision 3D Analysis Using 360° Electron Tomography. *Small Methods* **2018**, *2*, 1700276.

(36) Teich, E. G.; van Anders, G.; Klotsa, D.; Dshemuchadse, J.; Glotzer, S. C. Clusters of Polyhedra in Spherical Confinement. *Proc. Natl. Acad. Sci.* **2016**, *113*, E669–E678.

(37) Wang, D.; Hermes, M.; Kotni, R.; Wu, Y.; Tasios, N.; Liu, Y.; de Nijs, B.; van der Wee, E. B.; Murray, C. B.; Dijkstra, M.; van Blaaderen, A. Interplay between Spherical Confinement and Particle Shape on the Self-Assembly of Rounded Cubes. *Nat. Commun.* **2018**, *9*, 2228.

(38) Ducrot, É.; He, M.; Yi, G.-R.; Pine, D. J. Colloidal Alloys with Preassembled Clusters and Spheres. *Nat. Mater.* **2017**, *16*, 652–657.

(39) Ohfuji, H.; Rickard, D. Experimental Syntheses of Framboids - A Review. *Earth-Science Rev.* **2005**, *71*, 147–170.

(40) Ohfuji, H.; Akai, J. Icosahedral Domain Structure of Framboidal Pyrite. *Am. Mineral.* **2002**, *87*, 176–180.

(41) Damasceno, P. F.; Engel, M.; Glotzer, S. C. Predictive Self-Assembly of Polyhedra into Complex Structures. *Science* **2012**, *337*, 453–457.

(42) Yang, Y.; Wang, B.; Shen, X.; Yao, L.; Wang, L.; Chen, X.; Xie, S.; Li, T.; Hu, J.; Yang, D.; Dong, A. Scalable Assembly of Crystalline Binary Nanocrystal Superparticles and Their Enhanced Magnetic and Electrochemical Properties. *J. Am. Chem. Soc.* **2018**, *140*, 15038–15047.

(43) Kister, T.; Mravlak, M.; Schilling, T.; Kraus, T. Pressure-Controlled Formation of Crystalline, Janus, and Core–Shell Supraparticles. *Nanoscale* **2016**, *8*, 13377–13384.

(44) Bommineni, P. K.; Varela-Rosales, N. R.; Klement, M.; Engel, M. Complex Crystals from Size-Disperse Spheres. *Phys. Rev. Lett.* **2018**, *122*, 128005.

(45) Boles, M. A.; Engel, M.; Talapin, D. V. Self-Assembly of Colloidal Nanocrystals: From Intricate Structures to Functional Materials. *Chem. Rev.* **2016**, *116*, 11220–11289.

(46) Montanarella, F.; Geuchies, J. J.; Van Blaaderen, A.; Vanmaekelbergh, D. Crystallization of Nanocrystals in Spherical Confinement Probed by *In Situ* X-Ray Scattering. *Nano Lett.* **2018**, *18*, 3675–3681.

(47) Marino, E.; Kodger, T. E.; Wegdam, G. H.; Schall, P. Revealing Driving Forces in Quantum Dot Supercrystal Assembly. *Adv. Mater.* **2018**, *30*, 1–6.

(48) Galván-Moya, J. E.; Nelissen, K.; Peeters, F. M. Structural Ordering of Self-Assembled Clusters with Competing Interactions: Transition from Faceted to Spherical Clusters. *Langmuir* **2015**, *31*, 917–924.

(49) Sánchez-Iglesias, A.; Grzelczak, M.; Altantzis, T.; Goris, B.; Pérez-Juste, J.; Bals, S.; Van Tendeloo, G.; Donaldson, S. H.; Chmelka, B. F.; Israelachvili, J. N.; Liz-Marzán, L. M. Hydrophobic Interactions Modulate Self-Assembly of Nanoparticles. *ACS Nano* **2012**, *6*, 11059–11065.

(50) Sánchez-Iglesias, A.; Claes, N.; Solís, D. M.; Taboada, J. M.; Bals, S.; Liz-Marzán, L. M.; Grzelczak, M. Reversible Clustering of Gold Nanoparticles under Confinement. *Angew. Chemie - Int. Ed.* **2018**, *57*, 3183–3186.

(51) Pazos-Perez, N.; Wagner, C. S.; Romo-Herrera, J. M.; Liz-Marzán, L. M.; García De Abajo, F. J.; Wittemann, A.; Fery, A.; Alvarez-Puebla, R. A. Organized Plasmonic Clusters with High Coordination Number and Extraordinary Enhancement in Surface-Enhanced Raman Scattering (SERS). *Angew. Chemie - Int. Ed.* **2012**, *51*, 12688–12693.

(52) Vogel, N.; de Viguerie, L.; Jonas, U.; Weiss, C. K.; Landfester, K. Wafer-Scale Fabrication of Ordered Binary Colloidal Monolayers with Adjustable Stoichiometries.

*Adv. Funct. Mater.* **2011**, *21*, 3064–3073.

(53) Etienne, G.; Kessler, M.; Amstad, E. Influence of Fluorinated Surfactant Composition on the Stability of Emulsion Drops. *Macromol. Chem. Phys.* **2017**, *218*, 1–10.

(54) Biskupek, J.; Leschner, J.; Walther, P.; Kaiser, U. Optimization of STEM Tomography Acquisition - A Comparison of Convergent Beam and Parallel Beam STEM Tomography. *Ultramicroscopy* **2010**, *110*, 1231–1237.

(55) Hohmann-Marriott, M. F.; Sousa, A. A.; Azari, A. A.; Glushakova, S.; Zhang, G.; Zimmerberg, J.; Leapman, R. D. Nanoscale 3D Cellular Imaging by Axial Scanning Transmission Electron Tomography. *Nat. Methods* **2009**, *6*, 729–731.

(56) Gilbert, P. Iterative Methods for the Three-Dimensional Reconstruction of an Object from Projections. *J. Theor. Biol.* **1972**, *36*, 105–117.

(57) Bitzek, E.; Koskinen, P.; Gähler, F.; Moseler, M.; Gumbsch, P. Structural Relaxation Made Simple. *Phys. Rev. Lett.* **2006**, *97*, 170201.

(58) Anderson, J. A.; Lorenz, C. D.; Travesset, A. General Purpose Molecular Dynamics Simulations Fully Implemented on Graphics Processing Units. *J. Comput. Phys.* **2008**, *227*, 5342–5359.

(59) Frenkel, D.; Ladd, A. J. C. New Monte Carlo Method to Compute the Free Energy of Arbitrary Solids. Application to the Fcc and Hcp Phases of Hard Spheres. *J. Chem. Phys.* **1984**, *81*, 3188–3193.

(60) Schilling, T.; Schmid, F. Computing Absolute Free Energies of Disordered Structures by Molecular Simulation. *J. Chem. Phys.* **2009**, *131*.

For table of contents only

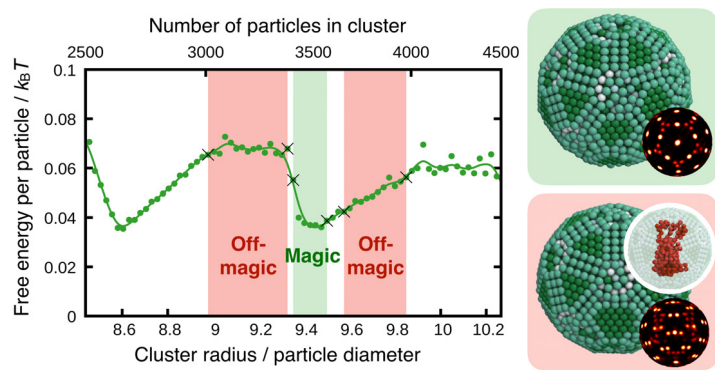